\documentclass[
    ,final            
    ,numberedheadings 
    ,4apaper          
    ,psfig
  ]
  {aipproc}

\layoutstyle{6x9}

\def\al{\alpha}

\def\ga{\gamma}
\def\de{\delta}

\def\ve{\varepsilon}

\def\et{\eta}
\def\th{\theta}

\def\la{\lambda}

\def\rh{\rho}

\def\si{\sigma}

\def\ta{\tau}

\def\La{\Lambda}

\def\Om{\Omega}

\def\cl{{\cal L}}

\def\fr#1#2{{{#1} \over {#2}}}

\def\frac#1#2{{\textstyle{{#1}\over {#2}}}}

\def\lsim{\mathrel{\rlap{\lower4pt\hbox{\hskip1pt$\sim$}}
    \raise1pt\hbox{$<$}}}
\def\gsim{\mathrel{\rlap{\lower4pt\hbox{\hskip1pt$\sim$}}
    \raise1pt\hbox{$>$}}}
\def\sqr#1#2{{\vcenter{\vbox{\hrule height.#2pt
         \hbox{\vrule width.#2pt height#1pt \kern#1pt
         \vrule width.#2pt}
         \hrule height.#2pt}}}}

\def\prt{\partial}
\def\lrpartial{\raise 1pt\hbox{$\stackrel\leftrightarrow\partial$}}

\newcommand{\beq}{\begin{equation}}
\newcommand{\eeq}{\end{equation}}
\newcommand{\bea}{\begin{eqnarray}}
\newcommand{\eea}{\end{eqnarray}}
\newcommand{\rf}[1]{(\ref{#1})}

\def\sech{\mathop{\rm sech}\nolimits}

\begin{document}

\title{Couplings varying on cosmological scales and Lorentz breaking}

\author{Ralf Lehnert}{
address={CENTRA, \'Area Departamental de F\'{\i}sica,
Universidade do Algarve, 8000-117 Faro, Portugal},
email={rlehnert@ualg.pt}
}

\begin{abstract}
In the context of $N=4$ supergravity in four dimensions, we
present an exact classical solution that leads to
spacetime-dependent electromagnetic couplings and discuss the
ensuing Lorentz-violating effects. We comment briefly on
experimental bounds.

\end{abstract}

\maketitle


\section{Introduction}


From a theoretical point of view, our current understanding of
nature at the fundamental level leaves unresolved a variety of
issues, so that present-day physical theories are believed to be
the low-energy limit of some underlying framework. Any effects
from an underlying theory involving gravity are expected to be
minuscule due to the likely suppression by at least one power of
the Planck mass. In such a situation, it appears practical to
consider violations of symmetries that hold exactly in our present
fundamental laws, might be violated in approaches to underlying
physics, and are amenable to high-precision experiments.

Although spacetime symmetries are a cornerstone of all known
physics, they might be violated at a more fundamental level: in
the context of string field theory, an explicit mechanism for the
spontaneous breaking of Lorentz invariance exists \cite{kps}.
Other examples of Lorentz-violating frameworks include
spacetime foam \cite{foam},
nontrivial spacetime topology \cite{klink},
realistic noncommutative field
theories \cite{ncqed}, and loop quantum gravity \cite{lgr}.
Moreover, Lorentz tests are currently among the most precise null
experiments available. Thus, spacetime-symmetry investigations
provide a promising tool in the search for underlying physics
\cite{cpt01}.

The low-energy effects of Lorentz breaking are described by a
general Standard-Model Extension \cite{ck}, which has been
constructed to contain all coordinate-invariant lagrangian terms
formed by
combining conventional field operators and coefficients carrying
Lorentz indices. Although these terms are observer Lorentz
symmetric, they explicitly break invariance under boosts and
rotations of particles \cite{rl03}. The Standard-Model Extension
has provided
the theoretical framework for the analysis of numerous
Lorentz-symmetry tests involving hadrons
\cite{hadronexpt,hadronth}, protons and neutrons \cite{pn},
leptons \cite{eexpt,eexpt2,ilt}, photons \cite{cfj,km,lipa,split},
muons \cite{muons}, and neutrinos \cite{neu}.

In the present work, we investigate the relation between Lorentz
breaking and violations of translation invariance \cite{sugra}.
More specifically, we argue that scalar couplings varying on
cosmological scales also lead to the type of Lorentz violation
described by the Standard-Model Extension. Since both Lorentz
transformations and spacetime translations are interwoven in the
Poincar\'e group, such a result does not come as a surprise.
Intuitively, the behavior of the vacuum is that of a
spacetime-varying medium so that isotropy, for example, can be
lost in certain local inertial frames.

Early work in the field of spacetime-dependent couplings includes
Dirac's large-number hypothesis \cite{lnh}. More recently, it has
been realized that varying couplings are natural in many
fundamental theories \cite{theo,recent}, which provides an additional
example for the fact that spacetime-symmetry violations are a
promising candidate experimental signature for more fundamental
physics. Investigations in this field are further motivated by
current claims of observational evidence for a time-varying
electromagnetic coupling \cite{webb}. The experimental status and
theoretical ideas are reviewed in Ref.\ \cite{jp}.

In the context of $N=4$ supergravity in four dimensions, we
demonstrate how smoothly varying couplings can naturally be
obtained from a classical cosmological solution. In particular,
the fine-structure parameter $\al=e^2/4\pi$ and
the electromagnetic $\th$ angle
acquire related spacetime dependences leading to the
aforementioned Lorentz-violating effects. Although the employed
supergravity framework is known to be unrealistic in detail, it is
contained in the $N=1$ supergravity in 11 spacetime dimensions,
which is a limit of M theory. Our approach can therefore yield
some insight into generic features of a candidate fundamental
theory.

This talk is organized as follows. In Sec.\ 2, we set up our
supergravity model in the context of cosmology. In particular, we
obtain an analytical solution to the equations of motion. Section
3 discusses the emergent time-varying couplings and comments on
experimental constraints. Aspects of the associated Lorentz
violation are investigated in Sec.\ 4. A short summary is
contained in Sec.\ 5.

\section{Supergravity cosmology}

The $N=4$ supergravity in four spacetime dimensions contains in
its spectrum a simple graviton represented by the metric
$g_{\mu\nu}$, four gravitinos $\psi^j_{\mu}$, six abelian
graviphotons $A_\mu^{jk}$, four fermions $\chi^j$, and a complex
scalar $Z$.
Latin indices $j,k,\ldots$ transform under the internal SO(4)
symmetry group, and the $A_\mu^{jk}$ lie in the adjoint
representation. In Planck units, the bosonic part $\cl$ of the
lagrangian takes the form \cite{cj} \beq \cl =
\sqrt{g}\left(-\frac 1 2  R -\frac 1 4 M_{jklm} F_{\mu\nu}^{jk}
F^{lm\mu\nu} -\frac 1 8  N_{jklm}\ve^{\mu\nu\rh\si}
F_{\mu\nu}^{jk} F^{lm}_{\rh\si} +  \fr {\prt_\mu Z
\prt^\mu\overline{Z}} {(1 - Z\overline{Z})^2}\right) . \label{lag}
\eeq The complex scalar $Z$ determines the generalized
electromagnetic coupling constant $M_{jklm}$ and the generalized
$\th$-term coupling $N_{jklm}$, which are both real: \beq M_{jklm}
+ i N_{jklm} = \frac 1 2(\de_{jl}\de_{km}-\de_{jm}\de_{kl})
\fr{1-Z^2}{1+Z^2} - i\ve_{jklm}\fr{ Z } {1+Z^2} \; .
\label{mplusn} \eeq Note also that $Z$ contains an axion and a
dilaton. It is convenient to isolate the dilaton piece $B$ via
a field
redefinition. Employing the Cayley map $W = -i(Z-1)/(Z+1)$ and
defining real fields $A$ and $B$ such that $W=A+iB$ yields the
following expression for the scalar kinetic term: $\cl_{\rm b}=
\sqrt{g} (\prt_\mu A\prt^\mu A + \prt_\mu B\prt^\mu B)/4B^2$. The
couplings $M_{jklm}$ and $N_{jklm}$ transform accordingly. The
fermion kinetic terms are just \beq \cl_{\rm
fermion}=\sqrt{g}\de^{jk}
\left(\overline{\psi}^\mu_j\ga_{\mu\nu\rh}D^\nu\psi^\rh_k+
\overline{\chi}_j\ga_{\mu}D^\mu\chi_k\right) . \label{fermions}
\eeq Note that these are independent of the scalars $A$ and $B$.
There are also higher-order terms in the fermions coupled to the
gauge fields and pieces that are quartic in the fermions.

In what follows, we look at situations in which only one
graviphoton, $F^{12}_{\mu\nu}\equiv F_{\mu\nu}$, is excited. The
bosonic lagrangian then takes the form \beq \cl
=\sqrt{g}\left(-\frac 1 2  R -\frac 1 4  M F_{\mu\nu} F^{\mu\nu}
-\frac 1 4  N F_{\mu\nu} \tilde{F}^{\mu\nu} + \fr{\prt_\mu
A\prt^\mu A + \prt_\mu B\prt^\mu B}{4B^2}\right), \label{lag2}
\eeq where we have abbreviated
$\tilde{F}^{\mu\nu}=\ve^{\mu\nu\rh\si}F_{\rh\si}/2$, as usual. The
electromagnetic and $\th$-term couplings become \beq M = \fr {B
(A^2 + B^2 + 1)} {(1+A^2 + B^2)^2 - 4 A^2}\; , \qquad N = \fr {A
(A^2 + B^2 - 1)} {(1+A^2 + B^2)^2 - 4 A^2}\; . \label{N} \eeq

Next, we construct an exact classical solution within the model
described by lagrangian \rf{lag2}. To this end, we consider a
homogeneous and isotropic Universe, with a flat ($k=0$)
Friedmann-Robertson-Walker (FRW) line element given by $ds^2 =
dt^2 - a^2(t) (dx^2 + dy^2 + dz^2)$. With these assumptions, the
scalars $A$ and $B$ and the scale factor $a$ can only depend on
the comoving time $t$. In the absence of energy-momentum sources
other than the scalar fields, one of the Einstein equations reads
$a\ddot{a}+2{\dot{a}}^2=0$. Besides the trivial solution $a=\rm
const.$, this equation is solved by \beq a(t)=c\sqrt[3\ ]{t}\: ,
\label{unreal} \eeq where $c$ is an integration constant. This
time evolution of the scale factor is far slower than the observed
one. This is a consequence of the fact that the above approach
fails to model the matter content of the Universe.

To describe a more realistic situation we refine our model by
including the energy-momentum tensor of dust given by $T_{\mu\nu}
= \rh u_\mu u_\nu$. Here, $u^\mu$ is a unit timelike vector
orthogonal to the spatial hypersurfaces and $\rh(t)$ is the
average energy density of galaxies and other matter. In the
present context, this energy-momentum tensor is associated with
the fermions in our model. Note that the fermionic sector does not
couple directly to the scalar fields $A$ and $B$, so that we take
$T_{\mu\nu}$ as conserved separately: \beq \fr{d(\rh a^3)}{dt} =
0\: . \label{cons} \eeq

For the moment, we set $F_{\mu\nu}$ to zero and consider the
equations of motion for our model. Variation of the action with
respect to $A$ and $B$ yields: \beq \fr d {dt} \left( \fr {a^3
\dot{A}}{B^2}\right) =0\:, \quad \fr d {dt} \left( \fr {a^3 \dot
B}{B^2} \right) + \fr {a^3 }{B^3} (\dot A^2 + \dot B^2) =0\: ,
\label{eq4} \eeq where the dot indicates a derivative with respect
to the comoving time $t$. Our supergravity model is also governed
by the Einstein equations. Varying with respect to the metric and
incorporating the energy-momentum tensor of dust gives: \beq
G_{\mu\nu} = T_{\mu\nu} + \fr 1 {2B^2} ( \prt_\mu A\prt_\nu A +
\prt_\mu B\prt_\nu B) - \fr 1 {4B^2} g_{\mu\nu} ( \prt_\la
A\prt^\la A + \prt_\la B\prt^\la B )\: . \label{einst} \eeq In the
present context, the ten equations \rf{einst} contain only two independent
ones: \beq - 3 \fr {\ddot a}{a} = \frac 1 2 \rh + \fr 1 {2B^2}
(\dot A^2 + \dot B^2)\:, \qquad \fr {\ddot a}{a} + 2 \fr {\dot a^2
}{a^2} = \frac 1 2 \rh\: . \label{eq2} \eeq

The energy-conservation equation \rf{cons} yields $\rh(t) =
c_n/a^3(t)$. The integration constant $c_n$ describes the amount
of fermionic matter in our model. If the Universe has matter
density $\rh_n$ and scale size $a_n = a(t_n)$ at the present time
$t_n$, then $c_n$ obeys $c_n = \rh_n a_n^3$. This result can be
used to integrate the second one of the Einstein equations
\rf{eq2}, which determines the time evolution of the scale factor:
\beq a(t) = \sqrt[3\ ]{ \frac 3 4 c_n (t+t_0)^2 - c_1}\:\: ,
\label{at} \eeq where $c_1$ and $t_0$ are integration constants
with the following physical interpretations: $c_1$ controls the
amount of energy stored in the scalar fields $A$ and $B$, and
$t_0$ sets the value of the comoving time $t$ at the initial
singularity. Our choice is $t_0 = \sqrt{4c_1/3c_n}$, which
corresponds to $t=0$ when $a(t) = 0$. Note that in this refined
version of our supergravity model including the dust, the time
dependence of the scale factor is $a(t) \sim t^{2/3}$ at late
times $t\gg t_0$, as anticipated for a $k=0$ matter-dominated
Universe.

The equation of motion for $A$ in \rf{eq4} yields $\dot A = c_2
B^2/a^3$, where the integration constant $c_2$ has been
introduced. The complete solution can most easily be obtained in
terms of a parameter time $\ta$ defined by \beq t = t_0
\left(\coth \fr{\sqrt{3}}{4\ta} - 1\right). \label{paramtime} \eeq
Note that $\ta = 0$ at the initial singularity when $t=0$, and
$\ta$ increases when $t$ increases. In terms of this parametric
time, the fields $A$ and $B$ evolve according to \beq A = \pm\la
\tanh (\fr 1 \ta + c_3) + A_0\: , \quad B = \la \sech (\fr 1 \ta +
c_3)\: . \label{be} \eeq Here $\la\equiv \mp 4 c_1/\sqrt{3} c_2
t_0$, and $c_3$ and $A_0$ are integration constants. In the
remaining part of this work we take $c_3$ to be zero for
simplicity. One can then verify that at late times on a scale set
by $t_0$, the parameter time obeys $\ta \approx \sqrt{3} t/4t_0$.
This implies that the late-time values of $A$ and $B$ are
given by $\pm 4 \la t_0/\sqrt{3} t + A_0$
and $\la (1 - 8t_0^2/3t^2)$, respectively.
Thus,
the axion and the dilaton tend to constant values.
Note in
particular, that this feature occurs for the string-theory dilaton
$B$, despite the absence of a dilaton potential. This is basically
a consequence of energy conservation.

\section{Spacetime-varying couplings}

The next step is to allow small fluctuations of $F_{\mu\nu}$. For
the moment, we take the axion-dilaton background determined by
\rf{be} as nondynamical. Many experiments are confined to
spacetime regions small on cosmological scales. We will therefore
continue our analysis in a local inertial frame.

The values of the couplings associated with the dynamics of the
field $F_{\mu\nu}$ are most easily extracted by comparison with
the conventional electrodynamics lagrangian in the presence of a
nontrivial $\th$ angle. In a local inertial frame, this lagrangian
can be taken as \beq \cl_{\rm em} = -\fr{1}{4 e^2}
F_{\mu\nu}F^{\mu\nu} - \fr{\th}{16\pi^2} F_{\mu\nu}
\tilde{F}^{\mu\nu}. \label{em} \eeq Then, inspection shows that in
our supergravity model we can identify \beq e^2 \equiv 1/M(t)\: ,
\qquad \th \equiv 4\pi^2 N(t)\: . \label{couplings} \eeq Note that
$M$ and $N$ are functions of the comoving time $t$ via Eq.\ \rf{be}. It
follows that in an arbitrary local inertial frame, $e$ and $\th$
acquire related spacetime dependences.

We continue with a few considerations regarding experimental
estimates. Our simplifying assumption is $c_3=0$, as mentioned
before. Matching the asymptotic electromagnetic coupling as
determined from the background \rf{be} with the observed
present-day value yields the boundary condition
$e^2(t\rightarrow\infty)\simeq4\pi/137$. It follows that
$|A_0|\simeq 1$ and $\la\lsim 2\pi/137$. Within this restricted
range of parameters, we take \beq \la = {2\pi}/{137}\: , \qquad
A_0 = \sqrt{1-\la^2}\: . \label{parameters} \eeq This special case
simplifies the analysis further because it leads to a zero
asymptotic value for $\th$. Note, however, that the above values
are sufficiently general in the sense that the estimates
determined below remain valid or improve for other parameter choices in more
than 98\% of the allowed range.

In what follows, we can replace the time coordinates $t'$ in
comoving local inertial frames with the comoving time $t$ because
$t'$ and $t$ agree to first order. At late times $t\gg t_0$, one
can verify that $e^2 \sim 2 \la \mp 8 \la^2 t_0/\sqrt{3}t$ and
thus $\dot \al/\al \sim \pm 4 \la t_0/\sqrt{3}t^2$. Observational
constraints on $\dot \al/\al$ at late times, i.e., in a recent
cosmological epoch, have been determined through various analyses
of data from the Oklo fossil reactor \cite{oklo}. The bounds
obtained are roughly $|\dot \al/\al| \lsim 10^{-16}$ yr$^{-1}$. If
the present age of the Universe is taken to be $t_n\simeq 10^{10}$
yr, the Oklo constraint yields the estimate $t_0 \lsim 10^{6}$ yr,
which is consistent with our previous late-times assumption.

In our supergravity model, the variation of the electromagnetic
coupling $\al$ with time can be relatively complicated, and
qualitative features of this variation can depend on the
integration constants determining the background \cite{nwap}. A
sample time dependence of $\al$ is depicted in Fig.\ 1. The solid
line represents the relative variation of $\al$ for the case
$t_n/t_0 = 2000$ as a function of the fractional look-back time
$1-t/t_n$ to the initial singularity. To provide an approximate
match to the recently reported data favoring a varying
fine-structure parameter,\footnote{The data, also plotted in Fig.\
1, were obtained from measurements of high-redshift spectra over
periods of approximately $0.6t_n$ to $0.8t_n$ assuming $H_0=65$
km/s/Mpc and   $(\Om_m ,\Om_\La)=(0.3,0.7)$ \cite{webb}.} the
integration constants $\la$ and $A_0$ have been changed fractionally by parts
in $10^{3}$ from the values \rf{parameters}. Although these
choices of parameters have no overlap with the Oklo data set, they
lie within the constraints for a variation of $\th$ to be discussed
in the next section. Note that the solid line reflects both
nonlinearities and non-monotonic features of $\al(t)$.

\begin{figure}
  \includegraphics[height=.4\textheight]{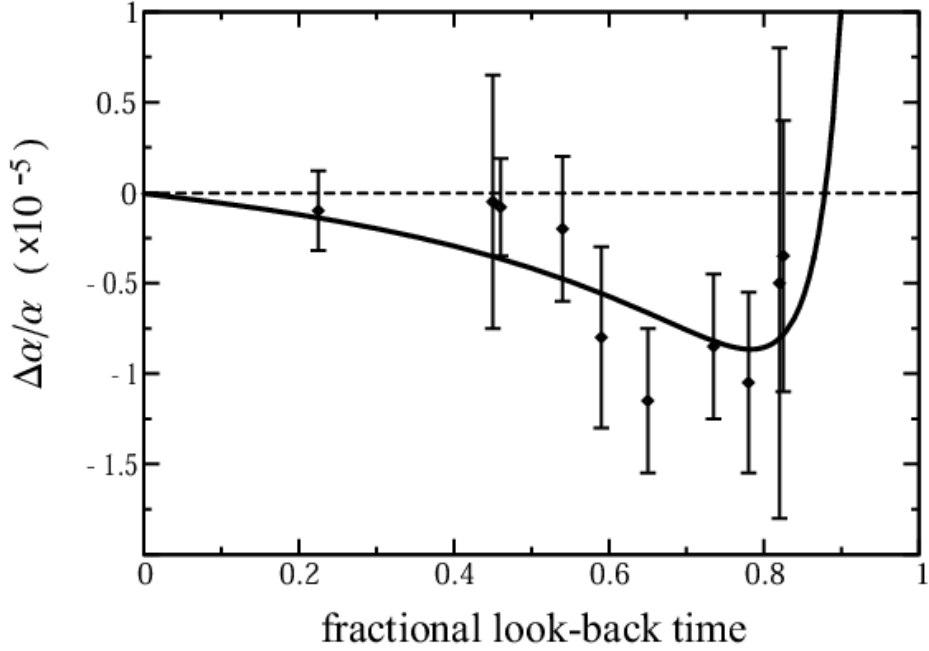}
  \caption{Sample relative variation of the fine-structure constant $\al$
    as function of the fractional look-back time $1-t/t_n$ to the Big Bang.
    The solid line corresponds to a parameter choice in the close vicinity
    of that given in \rf{parameters}.
    The dotted line represents a constant $\al$.
}
\end{figure}

\section{Lorentz violation}

The Lorentz-violating effects in our supergravity cosmology can be
seen explicitly at the level of the equations of motion for
$F_{\mu\nu}$: \beq \fr{1}{e^2}\partial_{\mu}F^{\mu\nu}
-\fr{2}{e^3}(\partial_{\mu}e)F^{\mu\nu}
+\fr{1}{4\pi^2}(\partial_{\mu}\th)\tilde{F}^{\mu\nu}=j^{\nu} .
\label{Feom} \eeq Here, we have introduced charged matter
described by a 4-current $j^{\nu}$ for completeness. When $e$ and
$\th$ are constant, their derivatives in \rf{Feom} would vanish
and the usual inhomogeneous Maxwell equations would emerge.
However, in the present case of varying $M$ and $N$, the dynamics
of the electromagnetic field is modified. Restricting attention to
spacetime regions small on cosmological scales, the gradients
$\prt_\mu M$ and $\prt_\mu N$ must be approximately constant.
Although these gradients do not spoil coordinate invariance, they
do select a preferred 4-direction in any local inertial frame. For
example, in a comoving inertial frame $\prt_\mu M$ and $\prt_\mu
N$ are both purely timelike. It follows that particle Lorentz
symmetry, i.e., symmetry under boosts, rotations, or both of
localized electromagnetic fields, is violated.

The above type of Lorentz breaking is to be distinguished from the
usual violation of global Lorentz symmetry in
textbook FRW cosmologies: in a conventional situation without
varying scalars any local inertial frame is Lorentz symmetric,
whereas in the present case the variation of $e$ and $\th$ results
in particle Lorentz breaking in all local inertial frames. Note
also that the Lorentz violation in our supergravity cosmology is
independent of the details of the model as long as $e$ and $\th$
vary on cosmological scales. This suggests particle Lorentz
violation could be a common feature of models incorporating
couplings with a sufficiently smooth and slow spacetime
dependence.

An integration by parts of the action yields an equivalent form of
our modified electrodynamics lagrangian: \beq \cl_{\rm
em}^{\prime} = -\fr{1}{4 e^2}  F_{\mu\nu} F^{\mu\nu}
+\fr{1}{8\pi^2}(\prt_\mu\th) A_\nu \tilde{F}^{\mu\nu} .
\label{prlagr} \eeq Since $\th$ varies in the present supergravity
model, particle Lorentz violation and CPT breaking are apparent
already at the lagrangian level. Again, in most practical
situations it suffices to consider small spacetime regions, so that
the gradient of $\th$ can be taken as a constant 4-vector. We can
then identify $e^2 \prt_\mu\th/8\pi^2$ with the Lorentz- and
CPT-violating $(k_{AF})_\mu$ parameter in the Standard-Model
Extension.

In addition to a constant $(k_{AF})_\mu$, consider now the special
situation in which $e$ does not vary. This case has recently
received a lot of attention in the literature
\cite{cfj,ck,jkk,neto,axph}. Then, the lagrangian \rf{prlagr} becomes
translationally invariant and energy-momentum conservation holds.
Note, however, that the conserved energy fails to be positive
definite, so that instabilities can occur \cite{cfj,ck,lehcpt01}.
On the other hand, the lagrangian \rf{prlagr} is associated with a
positive-definite supergravity theory\footnote{The $N$ term in the
lagrangian \rf{lag2} is independent of the metric and so does not
contribute to the conserved symmetric energy-momentum tensor. The
remaining terms have conventional structure and it is
straightforward to verify that they are positive definite.} and
the question arises how this difficulty is avoided in the present
context.

Although $(k_{AF})_\mu$ has been treated thus far as constant and
nondynamical, it is associated with the dynamical degrees of
freedom $A$ and $B$ in the present context. In the full theory,
excitations of the field $F_{\mu\nu}$ will lead to deformations
$\de A$ and $\de B$ in the background solution \rf{be}, such that
$A\rightarrow A+\de A$ and $B\rightarrow B+\de B$. Thus, in the
presence of a nonzero $F_{\mu\nu}$ the energy-momentum tensor
$(T^{\rm b})^{\mu\nu}$ of the background receives an additional
contribution from the perturbations $\de A$ and $\de B$, so that
$(T^{\rm b})^{\mu\nu}\rightarrow (T^{\rm b}_F)^{\mu\nu}= (T^{\rm
b})^{\mu\nu}+\de (T^{\rm b})^{\mu\nu}$. This contribution
compensates the negative-energy ones arising from the
$(k_{AF})_\mu$ term.

The compensation mechanism can be illustrated explicitly at the
classical level\footnote{In quantum field theory, radiative
corrections mix these terms \cite{fn2}.} in the lagrangian \beq
\cl=\cl_{\rm em}^{\prime}+\cl_{\rm b}\: , \label{lagcom} \eeq
where $\cl_{\rm b}$ has been defined in Sec. 2. In what follows we
concentrate on the $A$- and $B$-dependence of $\th$, and take as
$e$  as constant for simplicity. It can be checked that
incorporating the spacetime variation of $e$ leaves the
conclusions unchanged. We begin by considering the total conserved
energy-momentum tensor $(T^{\rm t}_F)^{\mu\nu}$ and isolating the
piece $(T^{\rm em})^{\mu\nu}$ associated with $F_{\mu\nu}$:
$(T^{\rm t}_F)^{\mu\nu}=(T^{\rm em})^{\mu\nu}+(T_F^{\rm
b})^{\mu\nu}$, where \bea (T^{\rm em})^{\mu\nu} & = & \fr{\prt
\cl}{\prt(\prt_{\mu}A^{\la})}\,\prt^{\nu}\! A^{\la}
-\et^{\mu\nu}\cl_{\rm em}^{\prime}\: ,\nonumber\\ (T_F^{\rm
b})^{\mu\nu} & = & \fr{\prt \cl}{\prt(\prt_{\mu}A)}\,\prt^{\nu}\!
A +\fr{\prt \cl}{\prt(\prt_{\mu}B)}\,\prt^{\nu}\! B
-\et^{\mu\nu}\cl_{\rm b}\: . \label{split} \eea With these
definitions we obtain explicitly: \beq (T^{\rm em})^{\mu\nu} =
\fr{1}{e^2}F^{\mu}_{\;\;\la}F^{\la\nu} +
\fr{1}{4e^2}\et^{\mu\nu}F^{\rh\si}F_{\rh\si}
+\fr{1}{8\pi^2}(\prt^\nu\th)A_{\la}\tilde{F}^{\la\mu} .
\label{emex} \eeq Note that only the last term in \rf{emex} can
lead to negative energies. Similarly, we find for the piece
associated with the background \beq (T_F^{\rm b})^{\mu\nu}  = \fr
{\prt^\mu A\prt^\nu A}{2B^2} -\fr {\et^{\mu\nu}} {4B^2} ( \prt_\la
A\prt^\la A + \prt_\la B\prt^\la B )
+\fr{\prt^\mu B\prt^\nu B}{2B^2}
-\fr{1}{8\pi^2}(\prt^\nu\th)A_{\la}\tilde{F}^{\la\mu}, \label{bex}
\eeq where again negative-energy contributions can arise only from
the last term. Equations \rf{emex} and \rf{bex} show that $(T^{\rm
t}_F)^{\mu\nu}$ is free from unsatisfactory terms, so that the
total conserved energy is positive definite, even in the presence
of a nonzero $(k_{AF})_\mu$. The apparent paradox lies in the fact
that the two pieces $(T_F^{\rm em})^{\mu\nu}$ and $(T_F^{\rm
b})^{\mu\nu}$, each containing the term with the positivity
difficulty, become separately conserved in the limit of a constant
$\prt^\nu\th$.\footnote{A constant timelike $(k_{AF})_\mu$
violates microscopic causality \cite{cfj,ck,klink,lehcpt01}. Our
supergravity model may circumvent this, but a complete analysis of
this lies outside our present scope.}

In the present supergravity cosmology, the spacetime dependences of
both $e$ and $\th$ follow from the background \rf{be} and are
therefore related. This fact can be exploited in the context of
experimental estimates. In our model, we obtain  $\dot{N} \sim \mp
2 t_0 / \sqrt{3} \la t^2$ for the time variation of $N$ at late
times. The direct observational limit of $(k_{AF})_0 \lsim
10^{-42}$ GeV \cite{cfj} then bounds the variation of $\al$ in
time to be $|\dot \al/\al| \lsim 10^{-12}$ yr$^{-1}$, consistent
with the Oklo data \cite{oklo}. Reversing the analysis, the Oklo
bounds constrain $(k_{AF})_\mu$ to be less than $\sim 10^{-46}$
GeV, which compares favorably with the above direct limit.

\section{Summary}

In a cosmological context, we have determined an analytical
solution within a simple supergravity model. This classical
solutions describes a situation with varying electromagnetic
couplings $e$ and $\th$. The functional dependence of these
couplings on spacetime is highly nonlinear. We have demonstrated
within this model and argued in the general case that
spacetime-dependent couplings lead to particle Lorentz violation.
Our supergravity cosmology avoids the usual positivity problems
associated with a varying $\th$ angle.


\begin{theacknowledgments}
The author wishes to thank V.\ Alan Kosteleck\'y and Malcolm J.\
Perry for their collaboration and Behram N.\ Kursunoglu for the
invitation to the Coral Gables meeting. This work was supported in
part by  the Centro Multidisciplinar de Astrof\'isica (CENTRA).

\end{theacknowledgments}





\end{document}